\documentstyle[prd,aps,epsfig]{revtex}
\begin{document}
\def\beqra{\begin{eqnarray}}
\def\eeqra{\end{eqnarray}}
\def\beq{\begin{equation}}
\def\eeq{\end{equation}}
\def\ds{\displaystyle}
\def\ts{\textstyle}
\def\ss{\scriptstyle}
\def\sss{\scriptscriptstyle}
\def\Vb{\bar{V}}
\def\phb{\bar{\phi}}
\def\rhb{\bar{\rho}}
\def\L{\Lambda}
\def\re#1{(\ref{#1})}
\def\D{\Delta}
\def\G{\Gamma}
\def\p{\partial}
\def\de{\delta}
\def \lta {\mathrel{\vcenter
     {\hbox{$<$}\nointerlineskip\hbox{$\sim$}}}}
\def \gta {\mathrel{\vcenter
     {\hbox{$>$}\nointerlineskip\hbox{$\sim$}}}}

\renewcommand{\Re}{\mathop{\mathrm{Re}}}{\renewcommand{\Im}{\mathop{\mathrm{Im}}}
\newcommand{\tr}{\mathop{\mathrm{tr}}}
\newcommand{\Tr}{\mathop{\mathrm{Tr}}}

%
%
\def\i{i}
\def\f{f}
\def\d{d}
\def\e{e}
\def\half{\mbox{\small $\frac{1}{2}$}}
\sloppy
\input epsf.sty 
\twocolumn

\draft
\title{{\Large {\bf Scalar-Tensor Gravity and Quintessence }}}
\author{{\large Nicola Bartolo$^1$ and Massimo Pietroni$^2$}}
\address{{\it $^1$ Dipartimento di Fisica, Universit\`a di Padova,}\\
Via F. Marzolo 8, I-35131 Padova, Italy\\
$^2$ INFN - Sezione di Padova,\\
Via F. Marzolo 8, I-35131 Padova, Italy}
\date{September 1, 1999}
\maketitle



\renewcommand{\Re}{\mathop{\mathrm{Re}}}{\renewcommand{\Im}{\mathop{%
\mathrm{Im}}} }

%
%

\abstract{\sl Scalar fields with inverse power-law effective potentials 
may provide a negative pressure component to the energy  density of the 
universe today, as required by cosmological observations. In order to
be cosmologically relevant today, the scalar field should have a mass $m_\phi 
=O(10^{-33} {\mathrm eV})$, thus potentially inducing sizable violations of 
the equivalence principle and space-time variations 
of the coupling constants. Scalar-tensor theories
of gravity provide a framework for accommodating phenomenologically acceptable 
ultra-light scalar fields. 
We discuss non-minimally coupled scalar-tensor theories in which
the scalar-matter coupling is a dynamical quantity.
Two attractor mechanisms are operative at the same time: 
one towards the tracker solution, which accounts
for the accelerated expansion of the Universe, and one towards general 
relativity, which makes the ultra-light scalar field phenomenologically safe 
today. As in usual tracker-field  models, the late-time behavior is largely 
independent on the initial conditions.
Strong distortions in the cosmic microwave background anisotropy spectra as
well as in the matter power spectrum are expected.
} 

DFPD/99/TH/40\\
\vskip 0.5cm
The growing  evidence for an accelerated expansion of the universe \cite
{scp,highz}, together with the analyses on cluster mass distributions \cite
{cluster} and preliminary data on the position of the first Doppler peak in
the cosmic microwave background anisotropies \cite{cmb}, strongly indicate
that a large component of the energy density of the universe has negative
pressure.

In a flat space-time, a cosmological constant $\Lambda $ with $\Omega _{\L
}\simeq 0.7$ could well play the role of this unknown component. Alternatively,
space-time dependent scalar fields have been considered, in what are usually 
called `quintessence models'. In particular,
models with inverse power-law effective potentials, $V(\Phi )=M^{4+m
}\Phi ^{-m }$, exhibit attractor solutions with negative pressure 
\cite{rp,liddle,zws,swz,mpr}, so that the presently observed accelerated
expansion of the universe might be ascribed to a scalar field which is still
rolling down its potential.
The good point about these models is the
existence of the attractor, which makes the present-time behavior nearly
independent on the initial conditions of the fields \cite{zws,swz}. On the
other hand, the field energy density scales with respect to that of the
background according to the power-law 
\begin{equation}
\frac{\rho _{\Phi }}{\rho _{B}}\sim a^{6\frac{1+w_{B}}{2+m }}\,,
\label{scal1}
\end{equation}
where the equation of state for the background is $w_{B}=0\;(1/3)$ for
matter (radiation). Then, assuming that the field reached the attractor
solution by the time of equivalence between matter and radiation ($
1+z_{eq}=2.4\cdot 10^4 \Omega_0 h_0^2 $) 
the above ratio has varied by a factor 
\[
z_{eq}^{6\frac{1+w_{B}}{2+m }} 
\]
since then. The explanation of why it is $\rho _{\Phi }\simeq \rho _{B}$
right today is the so called `cosmic coincidence' problem. For moderate
values of the exponent $m $ the required fine-tuning on the mass
parameter $M$ in the scalar effective potential is of the same order of that
needed in the case of a pure cosmological term ($m =0$ in eq. (\ref
{scal1})). Then, there is no clear improvement in this respect.

An even tougher problem emerges when the phenomenology of the scalar field $%
\Phi$ is taken into account. Indeed, the following relation holds on the
attractor solution \cite{zws} 
\[
V^{\prime\prime}=(9/2)(1-\omega_\Phi^2)[(m +1)/m] H^2\;, 
\]
which means that the scalar field is practically massless today, $m_\Phi
\sim H_0\simeq 10^{-33} {\mathrm eV}$. If the most general couplings of 
$\Phi$ with
the rest of the world are allowed, phenomenologically disastrous
consequences are induced, like violations of the equivalence principle and
time-dependence of gauge and gravitational constants on a time-scale $%
O(H_0^{-1})$ \cite{dam,car}. The former are strongly constrained by
E\"otv\"os type experiments to less than $10^{-12}$ level, whereas present
results on the time variation of coupling constants give $|\dot{\alpha}%
/\alpha| <6.7 \times 10^{-17} yr^{-1}$ for the electromagnetic coupling 
\cite{shly,dam2}, $|\dot{G}_F/G_F|< 10^{-12} yr^{-1}$ for the Fermi
constant \cite{shly} and $|\dot{G}/G|=(-0.2\pm 1.0) 10^{-11} yr^{-1}$ for
the Newton constant \cite{sha}.

Scalar-tensor theories of gravity (ST) represent a natural framework in
which massless scalars may appear in the gravitational sector of the theory
without being phenomenologically dangerous. In these theories the purely
metric coupling of matter with gravity is preserved, thus ensuring the
equivalence principle and the constancy of all non-gravitational coupling
constants \cite{dam}. Moreover, as discussed in \cite{dam3}, a large class
of these models exhibit an attractor mechanism towards general relativity
(GR), that is, the expansion of the Universe during the matter dominated era
tends to drive the scalar fields toward a state where the theory becomes
indistinguishable from GR.

In this letter we will discuss quintessence in the framework of ST theories.
We will identify a class of models in which two attractor mechanisms are
operative at the same time: one towards the tracker solution, which accounts
for the accelerated expansion of the Universe, and one towards GR, which
makes the ultra-light scalar field phenomenologically safe. In these models,
the coupling between the scalar field and ordinary matter is a dynamical
quantity which becomes smaller and smaller as the field rolls down its
effective potential. This is the main difference with respect to previous
works on ST theories as models for quintessence \cite{pbm,qst}, where the
attractor toward GR was not present and the coupling between the scalar
field and matter had to be fixed to small values once for all in order to
meet phenomenological constraints.

Moreover, we will find that during most of the matter-dominated era the
ratio $\rho_\Phi/\rho_m$ will scale with the {\it logarithm} of $a$ instead
of the power-law of eq. (\ref{scal1}). As a result, the variation of this
ratio from equivalence to today is sizably reduced, thus alleviating the 
coincidence problem.

ST theories of gravity are defined by the action \cite{dam,dam3} 
\begin{equation}
S=S_{g}+S_{m}\,,  \label{jordan0}
\end{equation}
where 
\begin{equation}
S_{g}=\frac{1}{16\pi }\int d^{4}x\sqrt{-\tilde{g}}\left[ \Phi^2 \tilde{R}-
4 \omega (\Phi ) \tilde{g}^{\mu \nu }\partial _{\mu }\Phi
\partial _{\nu }\Phi -2\tilde{V}(\Phi )\right] \,,  \label{jordan}
\end{equation}
and the matter fields $\Psi _{m}$ are coupled only to the metric tensor $%
\tilde{g}_{\mu \nu }$ and not to $\Phi $, {\it i.e.} $S_{m}=S_{m}[\Psi _{m},%
\tilde{g}_{\mu \nu }]$. $\tilde{R}$ is the Ricci scalar constructed from the
physical metric $\tilde{g}_{\mu \nu }$. Each ST model is identified by the
two functions $\omega (\Phi )$ and $\tilde{V}(\Phi )$. For instance, the
well-known Jordan-Fierz-Brans-Dicke (JFBD) theory \cite{bd} corresponds to $%
\omega (\Phi )=\omega $ (constant) and $\tilde{V}(\Phi )=0$. The matter
energy-momentum tensor, $\tilde{T}^{\mu \nu }=
2/\sqrt{-\tilde{g}}\,\delta S_{m}/\delta \tilde{g}_{\mu \nu },$ is conserved, 
masses and non-gravitational couplings are time
independent, and in a locally inertial frame non gravitational physics laws
take their usual form. Thus, the 'Jordan' frame variables 
$\tilde{g}_{\mu \nu }$
and $\Phi $ are also denoted as the 'physical' ones in the literature. On
the other hand, the equations of motion are rather cumbersome in this
frame, as they mix spin-2 and spin-0 excitations. A more convenient
formulation of the theory is obtained by defining two new gravitational
field variables, $g_{\mu \nu }$ and $\phi $ by means of the conformal
transformation 
\begin{eqnarray}
\tilde{g}_{\mu \nu } &\equiv &\displaystyle A^{2}(\phi )g_{\mu \nu } 
\nonumber \\
\Phi^2 &\equiv &\displaystyle A^{-2}(\phi )G_{\ast }^{-1}  \nonumber \\
V(\phi ) &\equiv &\displaystyle A^{4}(\phi )\tilde{V}(\Phi )  \nonumber \\
\alpha (\phi ) &\equiv &\displaystyle\frac{d\log A(\phi )}{d\phi }\,.
\end{eqnarray}
Imposing the condition 
\begin{equation}
\alpha ^{2}(\phi )=\frac{1}{2\omega (\Phi )+3}\,,
\end{equation}
the gravitational action in the `Einstein frame' reads 
\begin{equation}
S_{g}=\frac{1}{16\pi G_{\ast }}\int d^{4}x\sqrt{-{g}}\left[ {R}-2{g}^{\mu
\nu }\partial _{\mu }\phi \partial _{\nu }\phi -2{V}(\phi )\right] \,,
\end{equation}
and the matter one now contains also the scalar field $S_{m}[\Psi
_{m},A^{2}(\phi ){g}_{\mu \nu }]$. In this frame masses and
non-gravitational coupling constants are field-dependent, and the
energy-momentum tensor of matter fields is not conserved separately, but
only when summed with the scalar field one. On the other hand, the
gravitational constant $G_{\ast }$ is time-independent and the field
equations have the simple form 
\begin{eqnarray}
R_{\mu \nu }-\mbox{\small $\frac{1}{2}$}g_{\mu \nu }R &=&\displaystyle%
-g_{\mu \nu }(g^{\rho \sigma }\partial _{\rho }\phi \partial _{\sigma }\phi
+V(\phi ))+2\partial _{\mu }\phi \partial _{\nu }\phi  \nonumber \\
&&\displaystyle+8\pi G_{\ast }T_{\mu \nu }  \nonumber \\
&&  \nonumber \\
\displaystyle\partial ^{2}\phi -\mbox{\small $\frac{1}{2}$}\frac{\partial V}{%
\partial \phi } &=&\displaystyle-4\pi G_{\ast }\alpha (\phi )T\,,
\label{eomgen}
\end{eqnarray}
where $T_{\mu \nu }$ is the Einstein frame energy-momentum tensor, $T^{\mu
\nu }=2/\sqrt{-g}\,\delta S_{m}/\delta g_{\mu \nu }$.

The relevant point about the scalar field equation in (\ref{eomgen}) is that
its source is given by the trace of the matter energy-momentum tensor, $
T\equiv g^{\mu\nu}T_{\mu\nu}$, which implies the (weak) equivalence
principle. Moreover, when $\alpha(\phi)=0$ the scalar field is decoupled
from ordinary matter and the ST theory is indistinguishable from ordinary
GR. Indeed, at the post-newtonian level, the deviations from GR may be
parameterized in terms of an effective field-dependent newtonian constant%
\footnote{%
Strictly speaking, this is only true for a massless field, but for any
practical purpose it applies to our nearly massless scalars ($m_\phi\sim
H_0^{-1}$) as well.} 
\[
G=G(\phi)\equiv G_* A(\phi)^2 (1+\alpha^2(\phi))\,, 
\]
and two dimensionless parameters $\gamma$ and $\beta$ which, in the present
theories turn out to be \cite{dam} 
\begin{equation}
\gamma -1= -2\frac{\alpha^2}{1+\alpha^2}\,,\,\,\,\, \beta -1 = 
\mbox{\small
$\frac{1}{2}$} \frac{\kappa \alpha^2}{(1+\alpha^2)^2}\,,
\label{postnewt}
\end{equation}
where $\kappa=\partial\alpha/\partial \phi$.

The strongest bounds on the present values of the parameters $\alpha $ and $%
\kappa $ come from solar system measurements and may be summarized as
follows \cite{dam3} 
\begin{equation}
\alpha _{0}^{2}<10^{-3}\,,\,\,\,\,\,\,\,\,[(1+\kappa _{0})\alpha
_{0}^{2}]<2.5\times 10^{-3}\,.  \label{bounds}
\end{equation}

\bigskip

We next consider an homogeneous cosmological space-time 
\[
ds^{2}=-\ dt^{2}+a^{2}(t)\ dl^{2}\ , 
\]
where the energy-momentum tensor admits the perfect-fluid representation 
\[
T^{\mu \nu }=(\rho +p)\ u^{\mu }u^{\nu }\ +\ p\ g^{\mu \nu }\ \ , 
\]
with $g_{\mu \nu }\ u^{\mu }u^{\nu }=-1$.

The Friedmann-Robertson-Walker (FRW) equations then take the form 
\begin{eqnarray}
-3\frac{\ddot{a}}{a} &=&4\pi \ G_{\ast }\ (\rho +3\ p)+2\dot\phi^{2}-V(\phi ) 
\\
3\left( \frac{\dot{a}}{a}\right) ^{2}+3\frac{k}{a^{2}} &=&8\pi \ G_{\ast } 
\rho +
\dot\phi^{2}+V(\phi ) \\
\ddot{\phi} +3\frac{\dot{a}}{a}\dot\phi &=&-\ 4\pi \ G_{\ast }\ \alpha (\phi )
(\rho -3\ p)-
\frac{1}{2}\frac{dV(\phi )}{d\phi }\ ,
\end{eqnarray}
with the Bianchi identity 
\beq
d(\rho a^{3})+p\ da^{3}=(\rho -3\ p)\ a^{3}d\log A(\phi ). 
\label{bianchi}
\eeq

The physical proper time, scale factor, energy , and pressure, are related
to their Einstein frame counterparts by the relations 
\[
d\widetilde{\tau }=A(\phi )d\tau\,, \;\;\;\;\widetilde{a}=A(\phi)a\,,
\;\;\;\;
\widetilde{\rho }=A(\phi )^{-4}\rho\,,\;\;\;\; 
\widetilde{p}=A(\phi )^{-4}p. 
\]

\bigskip Defining new variables 
\[
\chi \equiv \log \frac{a}{a_0},\qquad \qquad 
\lambda \equiv \frac{V(\phi )}{8\pi G_{\ast
}\ \rho },\qquad \qquad w \equiv \frac{p}{\rho }, 
\]
and setting $k=0$ (flat space) the field equation of motion takes the more
convenient form \cite{dam3} 
\begin{eqnarray}
\ds \frac{2}{3-(\phi ^{\prime })^{2}}\ (1+\lambda )\ \phi ^{\prime \prime
}+[(1-w )+2\lambda ]\ \phi ^{\prime }=&&\nonumber\\
\ds -\alpha (\phi )\ (1-3\ w
)-\lambda \frac{d\log V(\phi )}{d\phi }, && \label{eom}
\end{eqnarray}
where primes denote derivation with respect to $\chi$.  
Eq. (\ref{eom}) will be our master equation.

Each ST theory is specified by a particular choice for $\alpha (\phi )$ and $%
V(\phi )$. As already mentioned, a constant $\alpha (\phi )=\alpha $ and $%
V=0 $ select the traditional JFBD theory, in which no mechanism of attraction
towards gravity is operative. In refs. \cite{pbm,qst} a non-zero potential with
power-law behavior was added to this case, and the model was studied
as a candidate for quintessence.

The mechanism of attraction towards GR can be illustrated by the simplest 
case $\alpha
(\phi )=\beta \ \phi $, which was studied in refs. \cite{dam3}. Choosing $V=0$
eq. (\ref{eom}) takes the form of the equation of motion of a particle with
velocity-dependent mass $m(\phi ^{\prime })=2/(3-(\phi ^{\prime })^{2})$ in
a parabolic potential $v=(1-3 w )\ \beta \ \phi ^{2}/2$, and subject to a
damping force proportional to $(1-w )$. Then, it is easy to realize that
at late times the field $\phi $ will settle down at the minimum of the
potential, $\phi =0$, where $\alpha (\phi )=0$ and the theory is
indistinguishable from GR.



In order to have a model for quintessence, the late-time behavior of the field
must be dominated by an effective potential with inverse-power law behavior. 
If this is the case, $\alpha(\phi)$ must decrease for large $\phi$, unlike the 
behavior considered above. We then consider the class of ST theories 
defined by
\beq
\alpha(\phi) = -B e^{-\beta \phi}\,,\;\;\;\;\;\;\;\;V(\phi) = D \phi^{-m}\,,
\eeq
 as models for quintessence.

From Bianchi identity we have the Einstein frame scaling laws
\[
\rho_{\mathrm rad} \sim a^{-4}\,\;\;\;\; \rho_{\mathrm mat} \sim A(\phi) 
a^{-3}\,,
\]
so that the background equation of state turns out to be
\beq
\ds w=\frac{1}{3} \frac{1}{1+\frac{A(\phi)}{A(\phi_{eq})} e^{\chi-\chi_{eq}}}\,
,
\eeq
$A(\phi_{eq})$ being the value of $A(\phi)$ at equivalence.

During radiation domination  $w \simeq 1/3$, 
thus one might expect the equation of motion 
(\ref{eom}) to be  insensitive to $\alpha(\phi)$. Actually, if one is 
interested 
in scalar fields with energy densities of the same order as matter today, then
the scalar potential term in the RHS of eq. (\ref{eom}) turns out to
be subdominant with respect to $\alpha(\phi) (1-3 w)$ during radiation 
domination and most of matter domination. In this regime, and neglecting
$\phi''$ with respect to $\phi^\prime$,
we find the approximate solution 
\beq
\phi(a) \simeq \frac{1}{\beta}\log\left[\beta B \log\left(\frac{2}{3}+
\frac{a}{a_{eq}}\right) + \mathrm{cost} \right]\,,
\label{appratt}
\eeq
which is an attractor in field space. In deriving eq. (\ref{appratt}) we have
also assumed $A(\phi) \simeq A(\phi_{eq})$ which is always the case after 
nucleosynthesis, if the bound (\ref{nucleobound}) below is satisfied.

Notice that this attractor is not the 
one  that would have been obtained in the $\alpha\,(1-3w) \to 0$ limit,
that is the well known tracker solution 
\beq
\phi_{\mathrm tr} \sim a^{\frac{3 (w_B+1)}{m+2}}\;, 
\label{trackrad}
\eeq
considered in  \cite{rp,liddle,zws,swz,mpr}, whose energy density scales 
according to eq. (\ref{scal1}).

The energy density of the solution (\ref{appratt}) is dominated by the
kinetic term ($w_\phi \simeq 1$) and scales according to
\beqra
\ds && \frac{\rho_\phi}{\rho_{B}} = 
\frac{1+\lambda}{3-(\phi^\prime)^2}(\phi^\prime)^2 + \lambda \nonumber \\
\ds && \simeq  \frac{1}{3} (\phi^\prime)^2 \sim 
\left( \frac{\frac{a}{a_{eq}}}{\frac{2}{3}+\frac{a}{a_{eq}}}\right)^2 
\left[
\beta B \log\left(\frac{2}{3}+
\frac{a}{a_{eq}}\right) + \mathrm{cost} \right]^{-2}\;.
\label{logge}
\eeqra

Thus, the field evolution may be schematically divided in two regimes;
at early times, during radiation domination and a large part of matter 
domination, it is  \mbox{ $|\alpha (1-3w)| \gg |\lambda \,d\log V/d\phi|$},
 and the relevant 
attractor is approximated by eqs. (\ref{appratt}), (\ref{logge}). 
Notice that after equivalence ($a>a_{eq}$) the field energy  varies 
only logarithmically with respect to the background energy so, in this regime,
the cosmic coincidence problem is parametrically alleviated.

At late times,
the potential term starts to dominate and the solution is attracted 
towards the usual tracker with negative pressure, 
eqs. (\ref{trackrad}), (\ref{scal1}), while the ST
theory flows towards GR as $\alpha \to 0$.

\begin{figure}[hb]
\begin{center}
\epsfig{file=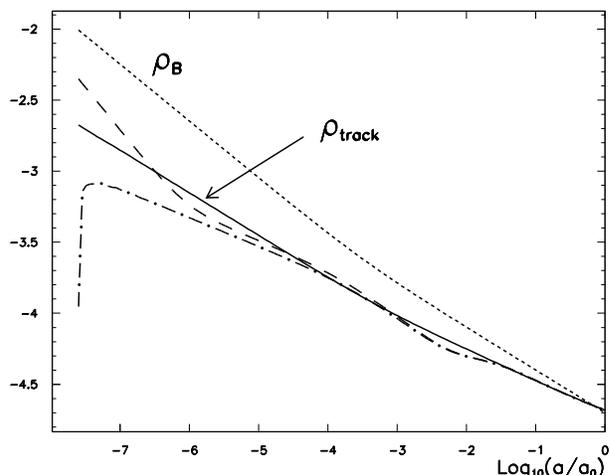,bbllx=30,bblly=200,bburx=560,bbury=600,height=6.3cm}
\caption{\scriptsize Energy densities (in $\log_{10^{10}}$ scale) 
{\it vs.} Einstein frame scale factor for $B=0.5$, $\beta=4$ and $m=6$. 
The 
short-dashed line is the background, whereas the long-dashed and 
dash-dotted are the field energy densities for two solutions with initial
energy much larger and smaller than that of the tracker (\ref{trackrad}) (solid
line), respectively.}
\end{center}
\end{figure}

In Fig. 1 the evolution of the energy density of typical solutions is shown as
a function of $\log_{10}(a/a_0)$.
The long-dashed and dash-dotted lines correspond to initial energies much 
larger and lower 
than the tracker solution (solid line), respectively. 
The latter has been obtained 
for  $\alpha=0$ and $\lambda \ll 1$ as in refs. 
\cite{rp,liddle,zws,swz}. The background energy density is plotted with the
dashed line. 

The potential term starts to dominate over the $\alpha$-term only at late times
$\log_{10}(a/a_0) \gta -1.5$. Before that epoch the two solutions have
already merged into the the attractor (\ref{appratt}). 
As we see, the good point about the tracker solutions of refs. 
\cite{rp,liddle,zws,swz}, namely the independence on the initial conditions,
is preserved in this ST model. Here, differently from 
\cite{rp,liddle,zws,swz,mpr} it is achieved by means of the attractor 
(\ref{appratt}) and not of the tracker (\ref{trackrad}).
Eq. (\ref{trackrad}) becomes the relevant attractor only for 
$\log_{10}(a/a_0) \gta -1.5$, when  the potential term starts to dominate
over the $\alpha$ one and the two solutions join the solid line.

\begin{figure}[hb]
\begin{center}
\epsfig{file=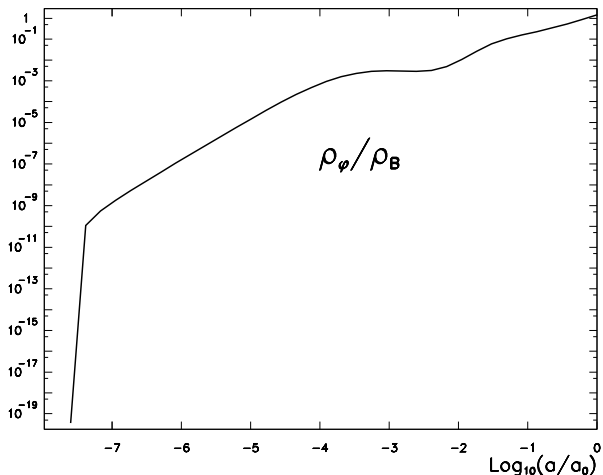,bbllx=30,bblly=200,bburx=560,bbury=600,height=6.3cm}
\caption{\scriptsize  Field to background energy ratio for the dash-dotted 
solution in Fig.1.}
\end{center}
\end{figure}

In Fig.2 we plot the field to background energy ratio for the dash-dotted
solution of Fig.1. Notice the logarithmic behavior from equivalence 
($\log_{10}(a/a_0) \simeq -4$) to the epoch in which
the potential term starts to dominate.

The total equation of state in the physical frame, 
$\widetilde{W}_{tot} \equiv \widetilde{p}_{tot}/\widetilde{\rho}_{tot}$, 
is plotted
in Fig. 3 versus the physical redshift $1+z \equiv \tilde{a}/\tilde{a_0}$. 

\begin{figure}[hb]
\begin{center}
\epsfig{file=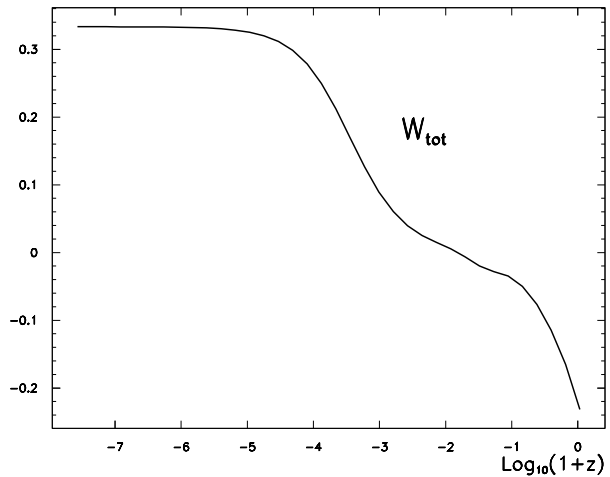,bbllx=30,bblly=200,bburx=560,bbury=600,height=6.3cm}
\caption{\scriptsize The total equation of state (matter+scalar field) in the
physical frame, for the dash-dotted solution in Fig.1.}
\end{center}
\end{figure}

Now we come to the phenomenological constraints on this ultra-light scalar.  
As we read from Fig. 4, the bounds coming from solar-system experiments, 
eq. (\ref{postnewt}) are largely satisfied by the present values of 
$\alpha^2$, so that we don't expect any deviation from GR to be measured
at present or in  forthcoming experiments. 

The strongest bounds come instead from nucleosynthesis. The variation of the 
(Jordan frame) Hubble parameter at nucleosynthesis 
induced by the time-dependent Newton constant
is given by
\beq
\frac{\Delta \tilde{H}^2}{\tilde{H}^2} = 1 - \frac{\Phi^2_{nuc}}{\Phi^2_{0}} =
1-\frac{A_0^2}{A_{nuc}^2}\;,
\eeq
and may be expressed in terms of the number of extra relativistic 
neutrino species as
\beq
\frac{\Delta \tilde{H}^2}{\tilde{H}^2}=\frac{7 \Delta N/4}{10.75+7 \Delta N/4}
\;.
\eeq
Taking the $95 \%$ CL limit for $\Delta N$, $\Delta N \le 1$ 
\cite{nucleo}
we get
\[
\frac{A_{0}^2}{A_{nuc}^2} \ge 0.86\;\;\;\;\;\;\;\;\;\; (95 \% \;\;\mathrm{CL})
\]
or, equivalently,
\beq \alpha_0-\alpha_{nuc} 
\le 0.08 \,\beta \;\;\;\;\;\;\;\;\;\; (95 \% \;\;\mathrm{CL}),
\label{nucleobound}
\eeq
which gives constraints on the ratio $B/\beta$ or on the value of the field at
nucleosynthesis. 

In Fig. 4 we plot $\alpha^2$ for the dash-dotted solution of Fig. 1. 
As we anticipated,
imposing the nucleosynthesis constrain (\ref{nucleobound}) the post-newtonian
bounds (\ref{postnewt}) turn out to be phenomenologically irrelevant.
On the other hand, strong signatures of the present scenario are generally 
expected
on  the anisotropy spectra of the Cosmic Microwave Background (CMB) as well 
as in the matter power-spectrum. 
These issues have been studied in ref. \cite{pbm} in the context of ST models
which do not exhibit the attractor behavior towards GR considered here.
The Induced Gravity (IG) model considered there is recovered  in our language 
with a field-independent $\alpha$, related to the parameter $\xi$ of 
ref. \cite{pbm} by
\[ 
\alpha^2_{IG} =\frac{2\xi}{1+6 \xi}\;.
\]
Being field independent, $\alpha^2_{IG}$ is bounded by (\ref{bounds}) at 
any epoch, and the resulting spectra distortions are at the some percent
level in IG \cite{pbm}.
On the other hand in the present case we may have $\alpha^2 = O(10^{-2})$ at 
decoupling, as we see in Fig. 4.   
Looking at Fig. 6 of ref. \cite{pbm} we can infer that such a large value
may decrease the height of the first
Doppler peak in the temperature anisotropy spectrum by a factor $\simeq 2$ 
and shift it to higher multipoles by $\Delta l \simeq 50$. 
Similar strong effects - well into the reach of future
MAP and PLANCK CMB experiments - are predicted for the CMB polarization 
anisotropy and the matter power spectra. 

\begin{figure}[hb]
\begin{center}
\epsfig{file=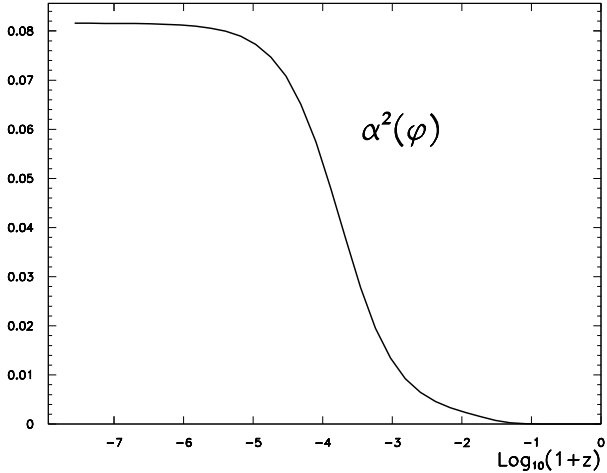,bbllx=30,bblly=200,bburx=560,bbury=600,height=6.3cm}
\caption{\scriptsize The dynamical evolution of the matter-scalar coupling 
$\alpha$ in the physical frame, for the dash-dotted solution in Fig.1.}
\end{center}
\end{figure}

\acknowledgments{It is a pleasure to thank Sabino Matarrese for many 
inspiring discussions.}

\end{document}